\documentstyle[prb,aps,multicol,epsf]{revtex}
\renewcommand{\narrowtext}{\begin{multicols}{2}
\global\columnwidth20.5pc}
\renewcommand{\widetext}{\end{multicols}
\global\columnwidth42.5pc} \multicolsep = 8pt plus 4pt minus 3pt

\begin{document}

\draft

\title{
Effect of in-plane magnetic field on magnetic phase transitions in
$\nu=2$ bilayer quantum Hall systems }

\author {Min-Fong Yang${}^1$ and Ming-Che Chang${}^2$}

\address{
${}^1$ Department of Physics, Tunghai University, Taichung, Taiwan
\\ ${}^2$ Department of Physics, National Taiwan Normal University,
Taipei, Taiwan }

\date{\today}
\maketitle

\begin{abstract}
By using the effective bosonic spin theory, which is recently proposed by Demler and
Das Sarma [ Phys. Rev. Lett. {\bf 82}, 3895 (1999) ], we analyze the effect of an
external in-plane magnetic field on the magnetic phase transitions of the bilayer
quantum Hall system at filling factor $\nu=2$. It is found that the quantum phase
diagram is modified by the in-plane magnetic field. Therefore, quantum phase
transitions can be induced simply by tilting the magnetic field. The general behavior
of the critical tilted angle for different layer separations and interlayer tunneling
amplitudes is shown. We find that the critical tilted angles being calculated agree
very well with the reported values. Moreover, a universal critical exponent for the
transition from the canted antiferromagnetic phase to the ferromagnetic phase is found
to be equal to $1/2$ within the present effective theory.


\end{abstract}

\narrowtext

Based on a microscopic Hartree-Fock theory and a long wavelength
field theory,\cite{Zheng} recent theoretical works predict that,
in the $\nu$=2 bilayer quantum Hall (QH) system under quite
general experimental conditions, there can be three qualitatively
different quantum magnetic phases: the fully spin polarized
ferromagnetic phase (F), the paramagnetic symmetric or spin
singlet (S) phase, and the intermediate canted antiferromagnetic
(C) phase. There have been some encouraging experimental evidence
for the existence of the novel C phase through inelastic light
scattering spectroscopy,\cite{Pellegrini1,Pellegrini2} transport
measurements,\cite{Sawada1,Sawada2} and capacitance
spectroscopy.\cite{Khrapai} However, it is difficult to make a
precise experimental verification of the proposed quantum phase
transitions, because a given sample (with fixed values of the
system parameters such as well widths, separations, etc.) is
always at a fixed point in the quantum phase diagram calculated in
Refs.~[\onlinecite{Zheng}]. Hence some extra parameters (for
example, the in-plane magnetic
field\cite{Pellegrini2,Sawada2,Khrapai} or the bias
voltage\cite{Sawada1,Sawada2,Brey}) are necessary in order to
allow experimentally a {\it continuous} tuning of the $\nu=2$ QH
system through the phase boundaries.

In this paper, we present a theoretical investigation of the
influence of an external in-plane magnetic field on the quantum
phase diagram of the bilayer system. Although the topology of the
phase diagram given by the Hartree-Fock theory should be correct,
the Hartree-Fock theory overestimates the stability of the C
phase.\cite{Zheng} In order to go beyond the Hartree-Fock theory,
an effective hard-core boson theory was proposed by Demler and Das
Sarma.\cite{Demler} Later, it is pointed out that the hard-core
bosons introduced in Ref. [\onlinecite{Demler}] form an exact
Schwinger boson representation of an effective spin-1/2 system,
and thus theoretical techniques developed for spin models can be
readily applied to the present problem.\cite{KYang1} Hence we
extend this effective model to the case that a parallel magnetic
field is applied, and study its effect on the magnetic phase
transitions of the bilayer QH system at $\nu=2$.\cite{note3} It is
found that the system may make multiple transitions from the S
phase to the C phase, and finally to the F phase as the parallel
magnetic field increases. The general behavior of the critical
tilted angle for different layer separations and interlayer
tunneling amplitudes is shown, and we find that the critical
tilted angles for the transition from the C phase to the F phase
coincide quantitatively with the experimental
values.\cite{Pellegrini2,Sawada2} Moreover, it is found that the
critical exponent of this transition has a universal value, which
is equal to $1/2$ within the present model.

In this work we assume that the electron-electron interaction and
Zeeman energies are much smaller than the Landau level splitting
due to the strong perpendicular magnetic field $B_\perp \hat{\bf
z}$, and we therefore restrict the orbital Hilbert space to the
lowest Landau level of either spin. When a parallel magnetic field
$B_\parallel$ along the $y$ direction is added, with the choice of
the gauge ${\bf A} = ( 0, B_\perp x, -B_\parallel x )$, the
Hamiltonian of the double-layer system has the form $H=H_0+H_{\rm
I}$ with\cite{note1}
\begin{eqnarray}
H_0&=&-\frac{{\bar \Delta}_{\rm SAS}}{2} \sum_{k,\sigma} \left( e^{-iQ k
l^2}c^\dagger_{1,k,\sigma} c_{2,k,\sigma} + h.c. \right) \nonumber\\ &&- \frac{{\bar
\Delta}_z}{2} \sum_{a,k,\sigma} \sigma c^\dagger_{a,k,\sigma}c_{a,k,\sigma}
\end{eqnarray}
and
\begin{eqnarray}
H_{\rm I}&=&{1\over2}\sum_{a_1,a_2}\sum_{k,p,q}
\sum_{\sigma_1,\sigma_2} V^{a_1 a_2}(q, k-p) \nonumber\\ && \quad
\times c^\dagger_{a_1,k+q,\sigma_1} c^\dagger_{a_2,p,\sigma_2}
c_{a_2,p+q,\sigma_2} c_{a_1,k,\sigma_1},
\end{eqnarray}
where $c _{a,k,\sigma }^{\dagger}$ creates an electron in the $a$-th layer ($a=1,2$)
with spatial  wavefunction $u_k(x,y) =
 e^{-iky} e^{-(x-k l^2)^2/2 l^2}/\sqrt{\pi^{1/2}l L}$ and
spin $\sigma / 2$ ($\sigma=\pm1$) in the direction of the {\it total} magnetic field
${\bf B} = B_\parallel \hat{\bf y} + B_\perp \hat{\bf z}$. ($L$ is the length of
system along the $y$ direction and $l=\sqrt{\hbar c/eB_\perp}$ is the magnetic
length.) Here the parameters in $H_0$ are defined as\cite{Hu}
\begin{eqnarray}
{\bar \Delta}_{\rm  SAS}&=& \Delta_{\rm  SAS} \exp(-Q^2 l^2 /4),
\label{new1}\\ {\bar \Delta}_z&=&\Delta_z \sqrt{ 1 + (
B_\parallel/B_\perp )^2}, \label{new2}
\end{eqnarray}
where $\Delta_{\rm SAS}$ is the tunneling-induced symmetric-antisymmetric energy
separation, and $\Delta_z$ is the Zeeman energy {\it in the absence of} $B_\parallel$.
We notice that $B_\parallel$ induces an Aharonov-Bohm phase factor $\exp(\pm iQ k
l^2)$ depending on the sense of interlayer tunneling with $Q=B_\parallel d/B_\perp
l^2$.

The matrix elements of the intralayer Coulomb interaction are
\begin{eqnarray}
V_A(p_1,p_2) &=& V^{11}(p_1,p_2) = V^{22}(p_1,p_2) \nonumber \\ &=&\sum_{\bf q} v_A(q)
\delta_{p_1,q_y} e^{-q^2 l^2/2} e^{iq_x p_2 l^2},
\end{eqnarray}
and the matrix elements of the interlayer Coulomb interaction are
\begin{eqnarray}
V_E(p_1,p_2) &=& V^{12}(p_1,p_2) = V^{21}(p_1,p_2) \nonumber \\ &=&\sum_{\bf q} v_E(q)
\delta_{p_1,q_y} e^{-q^2 l^2/2} e^{iq_x p_2 l^2}.
\end{eqnarray}
Here $v_A(q)=(2\pi e^2/\epsilon q)F_A(q,b)$ and $v_E(q)=v_A(q)F_E(q,b)e^{-qd}$ are the
Fourier transforms of the intralayer and the interlayer Coulomb interaction
potentials, respectively. $\epsilon$ is the dielectric constant of the system, and $d$
is the interlayer separation. In order to make a detailed comparison between the
experimental data and the theoretical predictions, we have also included
finite-well-thickness correction by introducing the form factor $F_A(q,b)$
($F_E(q,b)$) in the intralayer (interlayer) Coulomb matrix elements, where
$F_A(q,b)=2/bq-2(1-e^{-qb})/b^2q^2$, $F_E(q,b)=4\sinh^2(qb/2)/b^2q^2$, and $b$ is the
width of a quantum well.\cite{form}

Due to the Aharonov-Bohm phase factor in the tunneling process, if
the electrons of the bilayer system are in the symmetric state for
$B_\parallel=0$, the inclusion of an in-plane field twists the
original interlayer phase coherence and results in an increase in
the interlayer Coulomb energy.\cite{KYang2,YangSu} This effect can
be made explicit, if one absorbs the Aharonov-Bohm phase factor in
$H_0$ and makes the matrix elements of $H_0$ real by redefining
${\bar c}_{1,k,\sigma}=\exp(iQ k l^2/2) c_{1,k,\sigma}$ and ${\bar
c}_{2,k,\sigma} =\exp(-iQ k l^2/2) c_{2,k,\sigma}$, which can be
considered as a kind of pseudospin rotation. In terms of the new
operators ${\bar c}_{a,k,\sigma}$, the matrix elements of the
intralayer and the interlayer Coulomb interactions become
\begin{eqnarray}
{\bar V}^{11}(p_1,p_2) &=& {\bar V}^{22}(p_1,p_2)=V_A(p_1,p_2),
\nonumber \\  {\bar V}^{12}(p_1,p_2)&=&\left({\bar
V}^{21}(p_1,p_2)\right)^*=V_E(p_1,p_2)e^{iQ p_1 l^2}. \label{newV}
\end{eqnarray}
Notice that the matrix elements of the {\it interlayer} Coulomb
interaction become functions of $B_\parallel$.\cite{YangSu}
Consequently, in term of the ${\bar c}_{a,k,\sigma}$ operators,
the microscopic Hamiltonian at $B_\parallel \neq 0$ can be
considered as that at $B_\parallel=0$ with the modified matrix
elements ${\bar \Delta}_{\rm  SAS}$, ${\bar \Delta}_z$, and ${\bar
V}^{a_1 a_2}$. From these discussions, one can readily extend the
effective hard-core boson theory to the $B_\parallel \neq 0$ case.
We first give a brief review of the model for the $B_\parallel=0$
case,\cite{Demler,KYang1} and then provide the explicit relations
between the matrix elements of the microscopic Hamiltonian and the
parameters in the effective hard-core boson theory, such that the
effect of an external in-plane magnetic field can be easily
incorporated.

The Hamiltonian of a simple bilayer lattice model to describe the physics of the
bilayer $\nu=2$ QH system at $B_\parallel=0$ may be written as\cite{Demler}
\begin{eqnarray}
{\cal H} &=& - \frac{\Delta_{\rm SAS}}{2} \sum_i (
c_{1i\sigma}^{\dagger} c_{2i\sigma} + c_{2i\sigma}^{\dagger}
c_{1i\sigma} ) \nonumber\\ &&-\Delta_z \sum_i ( S^z_{1i} +
S^z_{2i} ) \nonumber\\  && + \frac{\epsilon_c}{2} \sum_i \left(
(n_{1i}-1)^2 + (n_{2i}-1)^2 \right) \nonumber\\ &&- J
\sum_{\langle ij \rangle} ( {\bf S}_{1i} {\bf S}_{1j} + {\bf
S}_{2i} {\bf S}_{2j} ) \label{Hamiltonian}
\end{eqnarray}
where $i$ is the in-plane site (intra-Landau-level) index, and $\sigma$ is the spin
index. ${S}^a_{1i} = \sum_{\alpha \beta} c_{1i\alpha}^{\dagger} \left(
\sigma^a_{\alpha\beta}/2 \right) c_{1i\beta}$ and $n_{1i} = \sum_{\sigma}
c_{1i\sigma}^{\dagger} c_{1i\sigma}$ are spin and charge operators for layer $1$, with
analogous definitions for layer $2$. The effective Heisenberg coupling $J$ and the
local charging energy $\epsilon_c$ of this model can be estimated as follows. In order
to have the same magnon spectrum in the F phase, one must impose that\cite{Girvin}
$Ja^2/2 = A l^2$ with the lattice constant $a=\sqrt{2 \pi} l$ and $A=(1/4)\sum_k
v_A(k) |k|^2 l^2 \exp(-k^2 l^2/2).$ Thus
\begin{equation}
J=\frac{1}{4\pi}\sum_k v_A(k) |k|^2 l^2 e^{-k^2 l^2/2}.
\end{equation}
On the other hand, $\epsilon_c$ can be estimated from $H_{\rm I}$
under the Hartree-Fock approximation for the system in the F
phase, which is given by
\begin{equation}\label{ec}
\epsilon_c = \sum_k V_A(k,0) - \sum_k V_E(k,0).
\end{equation}

Under the simplifications in which the total charge fluctuations are left out and only
the lowest two energy states for a given Landau orbital are kept, the effective
bilayer lattice model can be further reduced to a hard-core boson theory,\cite{Demler}
and it leads to a phase diagram which is more precise than that given by the
Hartree-Fock theory and is actually exact within the reduced Hamiltonian.\cite{KYang1}
The phase boundary separating the F and the C phases and that separating the C and the
S phases can be written as
\begin{eqnarray}
\Delta_z &=& -E_v - J(1-\sin 2\theta), \label{phase1} \\ \Delta_z
&=&-E_v - J(1+\sin 2\theta), \label{phase2}
\end{eqnarray}
where the parameters $E_v$ and $\theta$ are\cite{Demler}
\begin{eqnarray}
&&E_v=\frac{\epsilon_c}{2}- \sqrt{\Delta_{\rm SAS}^2+ (\epsilon_c/2)^2} , \label{Ev}\\
&&\theta=\tan^{-1}\left[ \frac{\epsilon_c/2}{\Delta_{\rm SAS}+\sqrt{\Delta_{\rm
SAS}^2+ (\epsilon_c/2)^2}} \right]. \label{theta}
\end{eqnarray}

Now we turn to the case of the presence of an in-plane magnetic field. As mentioned
before, when $B_\parallel \neq 0$, one needs to replace $\Delta_{\rm  SAS}$ and
$\Delta_z$ by ${\bar \Delta}_{\rm SAS}$ and ${\bar \Delta}_z$, respectively. Moreover,
the matrix elements of the interlayer Coulomb interaction become ${\bar V}_E(p_1,
p_2)=V_E(p_1, p_2) \exp(\pm i p_1 Q l^2)$. Therefore, the local charging energy
becomes
\begin{eqnarray}
{\bar \epsilon}_c&=&\frac{e^2}{\epsilon l} \int_0^\infty dx
e^{-x^2/2} F_A(x/l,b) \nonumber \\ &&-\frac{e^2}{\epsilon l}
\int_0^\infty dx e^{-x^2/2} e^{-xd/l} F_E(x/l,b) J_0(Qxl)
\label{new3}
\end{eqnarray}
($J_0(x)$ is the Bessel function), which is an increasing function of
$B_\parallel$.\cite{YangSu}

\begin{figure*}[h]
\centerline{\epsfxsize=6cm \epsfysize=6cm \epsfbox{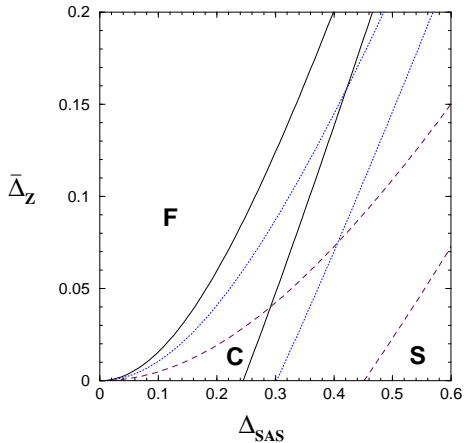} }
\caption{ $\nu$=2 bilayer phase diagrams in the bosonic spin
theory for different in-plane magnetic fields $B_\parallel$. Here
${\bar \Delta}_z$ means the Zeeman energy caused by the {\it
total} magnetic field. Continuous, dotted, and dashed lines
correspond to $B_\parallel/B_\perp=0, 1/\sqrt{3}$, and $1$,
respectively. The width of the electron layer is $1.0$ and the
interlayer separation is $1.45$. } \label{P_diag}
\end{figure*}

Substituting these modified parameters into Eqs.~(\ref{phase1})-(\ref{theta}), the
phase boundaries at nonzero $B_\parallel$ are obtained. The quantum phase diagram for
several in-plane magnetic fields is shown in Fig.~\ref{P_diag}. (In the followings,
the length and the energy units are chosen to be the magnetic length $l$ and the
intralayer Coulomb energy $e^2/\epsilon l$.) It is clear that the region of the F (S)
phase is expanded (shrunk) as $B_\parallel$ increases. The reason that the F phase is
enhanced and the S phase is suppressed comes from the fact that the system at
$B_\parallel \neq 0$ has a weaker ${\bar\Delta}_{\rm SAS}$ and a stronger
${\bar\Delta}_z$ compared to the system at $B_\parallel=0$. From this quantum phase
diagram, one finds that a parallel magnetic field can be a useful control parameter to
tune the system through the phase boundaries. For example, by applying $B_\parallel$,
the system can undergo a phase transition from the C phase to the F phase (or from the
S phase to the C phase). Indeed, such transitions are observed in recent tilted-field
experiments. \cite{Pellegrini2,Sawada2,Khrapai} The samples in
Refs.~[\onlinecite{Pellegrini2}] and [\onlinecite{Sawada2}] are initially in the C
phase near the F-C phase boundary in the absence of $B_\parallel$. The bilayer system
transits to the F phase when the tilted angle $\Theta=\tan^{-1}(B_\parallel/B_\perp)$
reaches a certain critical value $\Theta_C$. Thus the present theory can provide a
qualitative explanation of the observed transitions in these experiments.

\begin{figure*}[h]
\centerline{\epsfxsize=6cm \epsfbox{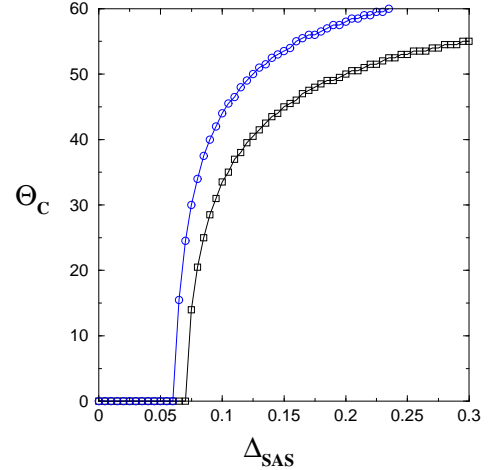} } \caption{
Critical tilted angles for the C-F phase transition are plotted as
a function of $\Delta_{\rm SAS}$. Circles are obtained using the
parameters for the sample in
Ref.[\protect\onlinecite{Pellegrini2}], where
$(\Delta_z,d,b)=(0.008,1.45,1.0)$; squares are for the sample in
Ref.[\protect\onlinecite{Sawada2}], where
$(\Delta_z,d,b)=(0.00687, 1.08,0.94)$. To compare the critical
tilted angles with experimental values, choose $\Delta_{\rm
SAS}=0.10$ for the first sample and $\Delta_{\rm SAS}=0.117$ for
the second sample. } \label{theta_1}
\end{figure*}

Moreover, we are able to compare the calculated $\Theta_C$ with
these experimental results quantitatively. Since it is relatively
easy to vary $\Delta_{\rm SAS}$ and $d$ in fabrication, we focus
our attention on the dependence of $\Theta_C$ on them. Our
theoretical prediction of the dependence of $\Theta_C$ on
$\Delta_{\rm SAS}$ (while the other parameters are fixed) is shown
in Fig.~\ref{theta_1}. We find that $\Theta_C$ is a monotonic
increasing function of $\Delta_{\rm SAS}$. It is expected because
the system with a larger $\Delta_{\rm SAS}$ locates at the phase
diagram with a farther distance to the F-C phase boundary, and a
larger bending of the F-C phase boundary caused by a bigger
$B_\parallel$ is necessary to make a transition to the F phase.
Adopting the sample parameters reported in
Refs.[\onlinecite{Pellegrini2}] and [\onlinecite{Sawada2}], we
obtain the theoretical values of $\Theta_C=33.5^\circ$ and
$48^\circ$, respectively (see Fig.~\ref{theta_1}), which agree
very well with the measured values: $\Theta_C \simeq 37^\circ$ in
Ref.[\onlinecite{Pellegrini2}] and $\Theta_C \simeq 50^\circ$ in
Ref.[\onlinecite{Sawada2}]. The close agreement gives us
confidence on the accuracy of the phase diagram calculated by
using the effective bosonic spin theory. We also find that, near
the transition point $\Delta_{\rm SAS}=\Delta_{\rm SAS}^{(c)}$,
both curves in Fig.~\ref{theta_1} (and other curves for several
sets of system parameters) fit very well with the function
$\Theta_c\simeq (\Delta_{\rm SAS}-\Delta_{\rm SAS}^{(c)})^\alpha$,
where $\alpha=1/2$. Indeed, by using series expansion for small
$\Theta_c$ and $\Delta_{\rm SAS}-\Delta_{\rm SAS}^{(c)}$, this
functional form can be derived from the formula of the F-C phase
boundary. This result means that $\alpha$ is a {\it universal}
critical exponent at the C-F phase transition, and its value
($\alpha=1/2$) may reflect the mean-field character of the present
approach.

\begin{figure*}[h]
\centerline{\epsfxsize=6cm \epsfysize=6cm  \epsfbox{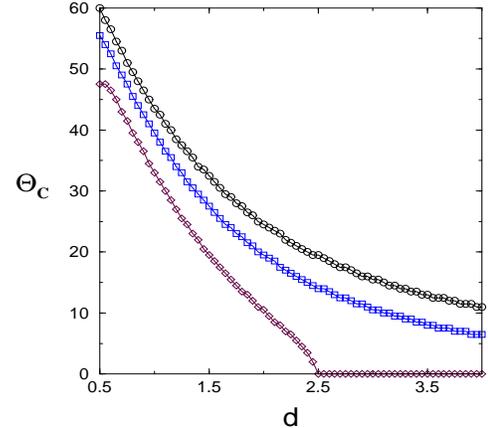} }
\caption{ Critical tilted angles for the C-F phase transition are
plotted as a function of layer separation $d$. All three curves
have the same $\Delta_z (=0.008)$ and $b$ (=1.0) The values of
$\Delta_{\rm SAS}$ for circles, squares, and diamonds are 0.1,
0.09, and 0.08 respectively. }  \label{theta_2}
\end{figure*}

The dependence of the critical tilted angle $\Theta_C$ on layer separation $d$ is
shown in Fig.~\ref{theta_2} for several different sets of sample parameters. It can be
seen that the critical tilted angle drops slowly at large $d$. In fact, for some cases
it seems that $\Theta_C$ may not drop to zero even when the layer separation
approaches infinity, which means that the C phase is stabilized in the single-layer
limit. This is an artifact due to fixing the interlayer tunneling amplitude
$\Delta_{\rm SAS}$ for a given curve. If the dependence of $\Delta_{\rm SAS}(d)$ can
be known and incorporated in the calculation, the critical tilted angles should
eventually drop to zero since the C phase is expected to disappear when the two layers
are completely decoupled. Therefore, the increase of $d$ in Fig.~\ref{theta_2} should
not be interpreted simply as a process of moving two layers apart. Rigorously
speaking, a single curve in Fig.~\ref{theta_2} is merely meant for samples with
different values of $d$ but {\it the same value} of $\Delta_{\rm SAS}$.

In conclusion, we have shown that the effective bosonic spin
theory with no adjustable parameters can describe the experimental
observations for the quantum phase transitions with quantitative
accuracy. Moreover, we find that the critical exponent of the C-F
phase transition has a universal value $\alpha=1/2$ within the
present effective model. It would be quite interesting to see
whether future experiments support our prediction or not. Finally,
the results presented here can be useful guidelines for
experimentalists to design their samples for observing the
magnetic phase transitions by tilting magnetic fields.

\acknowledgments The authors would like to thank H.~L.~Chiueh for
many valuable discussions. M.F.Y. also acknowledge financial
support by the National Science Council of Taiwan under contract
No. NSC 88-2112-M-182-003. M.C.C. is supported by the National
Science Council of Taiwan under contract No. NSC
88-2112-M-003-016.



\widetext


\begin{references}
\bibitem{note3}
The modification of the quantum phase diagram within the
Hartree-Fock theory can be easily obtained by using the formulas
of the phase boundaries in Ref.~[\protect\onlinecite{Zheng}] with
some suitable replacements of the matrix elements to take the
effect of the parallel magnetic field and the finite well
thickness into account (see the following discussions).
\bibitem{note1}
Here we assume that the two quantum wells are so narrow that the
effect of the in-plane magnetic field on the orbital degrees of
freedom can be ignored. The good agreement between our results and
the experimental observations may provide a justification of our
approach.
\end{references}
\end{document}